\documentclass[prd
,aps,twocolumn,showpacs,draft]{revtex4}

\begin{document}

\title{Essential on-Brane Equations for the Braneworld Gravity
\\ under the Schwarzschild Ansatz}
\author{			Keiichi Akama*}
\author{			Takashi Hattori**}
\author{			Hisamitsu Mukaida*}
\affiliation{	*Department of Physics, Saitama Medical University,
 			 Saitama, 350-0495, Japan}
\affiliation{	**Department of Physics, Kanagawa Dental College,
 			 Yokosuka, 238-8580, Japan}
\date{\today}

\begin{abstract}
It is argued that the braneworld gravity under the Schwarzschild ansatz 
	should obey three essential equations on the brane,
	and they are solved exactly.
We express the general solution of the fundamental equations of the braneworld 
	in power series of the brane normal coordinate.
The power series are derived by solving three independent 
	components of the bulk Einstein equation,
	and the coefficients are recursively determined 
	in terms of five functions on the brane.
The other components of the bulk Einstein equation are automatically satisfied, 
	as far as the five on-brane functions obey three essential equations.
They are the radial-extra and the extra-extra components
	of the bulk Einstein equation on the brane, 
	and the equation of motion of the brane.
Therefore the solution includes two arbitrary functions on the brane.
We show that the essential equations are exactly solved 
	by choosing an appropriate set of the two arbitrary functions.
The arbitrariness may affect the predictive powers 
	on the Newtonian and the post-Newtonian evidences.
\end{abstract}




\maketitle

\section{Introduction} 

Einstein gravity is successful in explaining 
	(i) the Newton's law of universal gravitation 
	for moderate distances, and 
	(ii) the post-Newtonian evidences (light deflections and
	planetary perihelion precessions due to solar gravity, etc.).
The explanations are achieved via the Schwarzschild solution 
	of the Einstein equation
	based on the ansatz 
	(a) staticity, 
	(b) spherical symmetry,
	(c) asymptotic flatness, and 
	(d) emptiness.
On the other hand, there are many plausible reasons that tempt us
	to take our 3+1 dimensional curved spacetime 
	as a ``braneworld" embedded in higher dimensions (``bulk")
	\cite{Fronsdal}--\cite{Akama:2011wi}.
There exist naive expectations that
	the braneworld theories inherit the successes of the Einstein gravity. 
It is, however, not a trivial problem because 
	they are not based on the brane Einstein equation, 
	but on the bulk Einstein equation and 
	the equation of motion of the braneworld.

In the previous paper, we obtained the general solution
	of the diagonal components of the equations at the brane,
	and found a large arbitrariness \cite{Akama:2010rb}.
In this paper, 
	we show that the braneworld gravity under the Schwarzschild ansatz 
	should obey three essential equations on the brane,
	and that they are exactly solvable.
We consider the general solution
	of the full set of the fundamental equations all over the bulk
	under the {Schwarzschild ansatz} (a)--(d) \cite{Akama:2011wi}.
For definiteness, we consider the Nambu-Goto brane
	interacting with the five dimensional bulk Einstein gravity.
The general solution is expressed in power series of the brane normal coordinate
	in terms of five functions on the brane.
The five functions should obey three on-brane equations,
	the radial-extra and the extra-extra components
	of the bulk Einstein equation on the brane, 
	and the Nambu-Goto equation.
They are the equations which determine how the brane is curved 
	according to the dynamics, 
	and, hence, they are the essential equations for the braneworld gravity.
The equations are solved
	exactly in terms of two arbitrary functions on the brane.
The arbitrariness may affect the predictive powers 
	on the Newtonian and the post-Newtonian evidences.

For the braneworlds, the {Schwarzschild ansatz} requires
	more precise specifications of the meaning.
We impose the staticity (a) all over the bulk, 
	while the asymptotic flatness (c) only on the braneworld,
	and not necessarily outside.
As for the sphericity (b), we assume symmetry under all the global rotations 
	which maps the braneworld onto itself.
We assume the emptiness (d) only outside the braneworld, 
	and allow matter and energy distributions 
	within a thin region inside the brane.
The empty region is not necessarily infinite, 
	but can be bounded due to topological structures, 
	other objects, or other branes.

In Sec.\ II, we specify the fundamental equations to be solved.
In Sec.\ III and IV, for the purpose of preparation, 
	we consider the bulk Einstein equation alone 
	(omitting the Nambu-Goto equation).
In Sec.\ V, including the Nambu-Goto equation,
	we show the three essential equations for the braneworld gravity,
	which are exactly solvable with two arbitrary functions.

\section{Fundamental Equations}

In order to seek for the general solution,
	we begin with examining what are the equations to be solved.
For definiteness, we consider the 3+1 dimensional Nambu-Goto type brane 
	in 4+1 curved spacetime with Einstein type gravity.
Let $X^I$ be the bulk coordinate, and
	$g_{IJ}(X^K)$ be the bulk metric 
	at the point $X^K$ \cite{notation}.
Let the brane be located at $X^I=Y^I(x^\mu)$ in the bulk, 
	where $x^\mu$ ($\mu=0,1,2,3$) are parameters
	which serve as the brane coordinate.
The dynamical variables of the system are 
	$g_{IJ}(X^K)$, $Y^I(x^\mu)$, and matter fields. 
The $Y^I(x^\mu)$ should be taken as the collective modes of the brane 
	formed by matter interactions.
Note that it is inappropriate to take the induced metric 
	$\tilde g_{\mu\nu}=Y^I_{,\mu} Y^J_{,\nu} g _{IJ}(Y^K)$
	as a dynamical variable,
	since it alone cannot completely specify the state of the braneworld
	\cite{notation}.
Then, the action integral is given by
\begin{eqnarray}
	\int\!\!\sqrt{- g}(\kappa^{-1} R-2\lambda)d^5 X
	-2\tilde\lambda \!\int\!\! \sqrt{-\tilde g } d^4 x+ S_{\rm m},
  \label{action}
\end{eqnarray}
where 	$ S_{\rm m}$ is the matter action,
	$g=\det g_{IJ}$, 
	$ R = R^I_{\ I}$,
	$ R_{IJ}= R^K_{\ IJK}$, 
	$ R^L_{\ IJK}$ 
	is the bulk curvature tensor
	written in terms of $ g _{IJ}$,
	$\tilde g =\det \tilde g_{\mu\nu} $, 
	and  $ \lambda $, $ \kappa$, and $\tilde \lambda$ 
	are constants.
We add no artificial fine-tuning terms.
The equations of motion are derived by 
	varying the action (\ref{action}) 
	with respect to the dynamical variables 
	$g^{IJ}$, $Y^I$ and the matter fields:
\begin{eqnarray}&&
	{\cal E}_{IJ}\equiv R_{IJ}-\frac{1}{2} R g _{IJ}
	+ \kappa (T_{IJ}+ \lambda g _{IJ})=0,
  \label{BE}
\\&&
	{\cal N}^{I}\equiv (\tilde \lambda \tilde g^{\mu\nu}
	+\tilde T^{\mu\nu})
	Y^I_{;\mu\nu}=0,\ \ \ \ \ \ 
  \label{NG}
\end{eqnarray}
and those for matters,
where 	$T_{IJ}$ and $\tilde T_{\mu\nu}$ are the energy momentum tensors 
	with respect to $g_{IJ}$ and $\tilde g_{\mu\nu}$, 
	respectively \cite{notation}.
Eq.\ (\ref{BE}) is the bulk Einstein equation, and 
	eq.\ (\ref{NG}) is the Nambu-Goto equation
	for the {braneworld dynamics}. 

In accordance with the ansatz of staticity and sphericity, 
	we introduce time, redial, polar and azimuth coordinates,
	$X^0=t$, $X^1=r$, $X^2=\theta$, and $ X^3=\varphi$, respectively,
	and the normal geodesic coordinate $X^4=z$, 
	such that $X^\mu=x^\mu$ ($\mu=0,\cdots,3$) and $z=0$ on the brane.
According to staticity and sphericity we can generally choose 
	the metric tenser $ g _{IJ}$ of the form 
\begin{eqnarray}&&
	ds^2= g _{IJ} dX^I dX^J
\cr&&
	=fdt^2-hdr^2-k(d\theta^2+\sin^2\theta d\varphi^2)-dz^2,
  \label{ds2}
\end{eqnarray}
where	$f$, $h$ and $k$ are functions of $r$ and $z$ only,
	and we choose as $k|_{z=0}=r^2$ using deffeomorphism.
Asymptotic flatness of the brane implies that 
\begin{eqnarray}&&
	f,\ h\rightarrow 1\ \ \ \
	{\rm as} \ \ \ \ r\rightarrow\infty\ \ \ \  
	{\rm at} \ \ \ \ z=0. 
  \label{f,h->1}
\end{eqnarray}
The independent non-vanishing components of the Ricci tensor $R_{IJ}$ are
\begin{eqnarray}&&
	R_{00}=- f_{zz}/2+ f_z{}^2/4f - f_z h_z/4h- f_z k_z/2k
\cr&&\ \ \ \ \ \ \ 
	- f_{rr}/2h+ f_r{}^2/4fh+ f_r h_r/4h^2- f_r k_r/2kh,\ \ \ \ \ 
\label{R00}
\\&&%
	R_{11}=h_{zz}/2-h_z{}^2/4h
	+{f_{z} h_{z}}/{4f}+ {h_{z} k_{z}}/{2k}
\cr&&\ \ \ \ \ \ \ \ \ 
	+ f_{rr}/2f- f_r{}^2/4f^2- f_r h_r/4fh\ \ \ \ 
\cr&&\ \ \ \ \ \ \ \ \ \ 
	+ k_{rr}/k- k_r{}^2/2k^2- k_r h_r/2kh,\ \ \ \ 
\label{R11}
\\&&%
	R_{22}= k_{zz}/2+ f_z k_z/4f+ h_z k_z/4h
\cr&&\ \ \ \ \ \ \ \ \ 
	+ k_{rr}/2h+ f_r k_r/4fh- h_r k_r/4h^2-1, \ \ \ \ 
\label{R22}
\\&&%
	R_{44}= f_{zz}/2f+ h_{zz}/2h+ k_{zz}/k
\cr&&\ \ \ \ \ \ \ \ \ 
	- f_z{}^2/4f^2- h_z{}^2/4h^2- k_z{}^2/2k^2, \ \ \ \ 
\label{R44}
\\&&%
	R_{14}= f_{zr}/2f + k_{zr}/k - f_z f_r/4f^2
\cr&&\ \ \ \ \ \ \ \ \ 
	- h_z f_r/4fh- h_z k_r/2hk - k_z k_r/2k^2, \ \ \ \ 
\label{R14}
\end{eqnarray}
where subscripts $r$ and $z$ indicate partial differentiations.
The ansatz of staticity and sphericity also indicate that 
	the only independent nonvanishing components 
	of the energy momentum tensors are
	$T_{00}$, $T_{11}$, $T_{22}$, $T_{14}$, $T_{44}$,
	$\tilde T_{00}$, $\tilde T_{11}$ and $\tilde T_{22}$.

\section{Off-Brane Solution of the Bulk Einstein Equation}

We first solve the bulk Einstein equation (\ref{BE}) alone
	without brane dynamics (\ref{NG}).
We have five equations for three functions $f$, $h$ and $k$.
The equations are not all independent
	because the conservation law holds:
\begin{eqnarray}
	g^{JK}{\cal E}_{IJ;K}=0. 
  \label{cons}
\end{eqnarray}
We rewrite the equation (\ref{BE}) into the equivalent form
\begin{eqnarray}
	{\cal R}_{IJ} \equiv R_{IJ} + \kappa 
	(T_{IJ} - T g _{IJ}/3 - 2\lambda g _{IJ}/3)=0, 
  \label{FIJ}
\end{eqnarray}
which is more convenient in deriving solutions below.
In terms of ${\cal R}_{IJ}$, we have ${\cal E}_{IJ}={\cal R}_{IJ}-{\cal R}g_{IJ}/2$.
For ${\cal R}_{IJ}$ with (\ref{R00})--(\ref{R14}), 
	the conservation law (\ref{cons}) is written as
\begin{eqnarray}&&\hskip-10pt
	{\cal R}_{14,4} = -{1 \over 2f} {\cal R}_{00,1}
	- {1 \over 2h} {\cal R}_{11,1}+{1 \over k} {\cal R}_{22,1}
	+{1 \over 2} {\cal R}_{44,1}
\cr&&\hskip-10pt 
	-\left(\frac{f_r}{2f}-\frac{h_r}{2h}+\frac{k_r}{k} \right)
	\frac{{\cal R}_{11}}{h}
	-\left(\frac{f_z}{2f}+\frac{h_z}{2h}+\frac{k_z}{k}\right) {\cal R}_{14},
	\ \ \ \ \ \ 
  \label{F14,4}
\\&& \hskip-10pt
	{\cal R}_{44,4} =-{1 \over f} {\cal R}_{00,4}
	+ {1 \over h} {\cal R}_{11,4}+{2 \over k} {\cal R}_{22,1}
	- {2\over h} {\cal R}_{14,1}
\cr&&\hskip-10pt 
	-\left(\frac{f_r}{f}-\frac{h_r}{h}+\frac{2k_r}{k}\right)
	\frac{{\cal R}_{14}}{h}
	-\left(\frac{f_z}{f}+\frac{h_z}{h}+\frac{2k_z}{k}\right)
	{\cal R}_{44}.\ \ \ \ \ \ 
  \label{F44,4}
\end{eqnarray}
We expand quantities in terms of $z$. 
We denote by ${F}^{[n]}(r)$ the coefficient of the $z^n$ term 
	in the expansion of any function ${ F} (r,z)$:
\begin{eqnarray}
	{F} (r,z)=\sum_{n=0}^{\infty}{F} ^{[n]}(r) z^n.
\label{calF=}
\end{eqnarray}
The operations of $^{[n]}$ obey the reduction rules 
\begin{eqnarray}&&
	({F}+{G})^{[n]}= {F}^{[ n]}+ {G}^{[ n]},\ \ \ 
	(c{F})^{[n]}= c{F}^{[ n]},\ \ \ 
\label{nred1}
\\&&
	({FG})^{[n]}= {\sum}_{k=0}^{n}{F}^{[ k]} {G}^{[ n- k]},\ \ \ 
\\&&
	{ F}^{-1}{}^{[n]}
	=-{\sum}_{k=0}^{n-1}{ F}^{-1}{}^{[ k]} { F}^{[n-k]}
	  { F}^{-1}{}^{[0]},
\label{nred3}
\end{eqnarray}
	where ${ F}$ and ${ G}$ are functions
	and $c$ is a constant. 
Let us assume ${\cal R}_{00}= {\cal R}_{11}= {\cal R}_{22}=0$.
Then, we have	
\begin{eqnarray}&& \hskip-10pt
	{\cal R}_{14}^{[n]} = \frac{1}{n} \bigg[{1\over 2}{\cal R}_{44,1}
	-\left(\frac{f_z}{2f}+\frac{h_z}{hf}+\frac{k_z}{k} \right) {\cal R}_{14}
	\bigg]^{[n-1]},\ \ 
  \label{F14n}
\\&& \hskip-10pt
	{\cal R}_{44}^{[n]} =\frac{1}{n}\bigg[- {2\over h} {\cal R}_{14,1}
	-\left(\frac{f_r}{f}-\frac{h_r}{h}+\frac{2k_r}{k}\right)
	\frac{{\cal R}_{14}}{h}
\cr&& \hskip-10pt
\ \ \ \ \ \ \ \ \ \ \ \ \ \ \ \ \ \ \ \ \ \ \ 
	-\left(\frac{f_z}{f}+\frac{h_z}{h}+\frac{2k_z}{k} \right) {\cal R}_{44}
	\bigg]^{[n-1]} \ \ \ \ \ \ \ \ 
  \label{F44n}
\end{eqnarray}
for $n\ge1$. 
According to (\ref{nred1})--(\ref{nred3}),
the right-hand side of eqs. (\ref{F14n})--(\ref{F44n})
	are linear combinations of ${\cal R}_{14}^{[j]}$ and ${\cal R}_{44}^{[j]}$
	with $0\le j \le n-1$ and their $r$-derivatives.
Therefore, if we have ${\cal R}_{14}^{[0]}= {\cal R}_{44}^{[0]}=0$, 
	we can conclude that ${\cal R}_{14}^{[n]}= {\cal R}_{44}^{[n]}=0$
	for any $n\ge1$.
Thus, the independent equations to be solved are 
\begin{eqnarray}&&
	{\cal R}_{00}= {\cal R}_{11}= {\cal R}_{22}=0 
	\ \ \ {\rm in\ the\ bulk\ and}
	\label{F00=F11=F22=0}
\\&& 
	{\cal R}_{14}= {\cal R}_{44}=0 
	\ \ \ \ \ \ \ \ \ \ \ \ {\rm on\ the\ brane.}  
	\label{F140=F440=0}
\end{eqnarray}
If we keep (\ref{F00=F11=F22=0}), we can replace (\ref{F140=F440=0}) by 
\begin{eqnarray}&&
	{\cal E}_{14}= {\cal E}_{44}=0 
	\ \ \ \ \ \ \ \ \ \ \ \ {\rm on\ the\ brane.}  
	\label{E140=E440=0}
\end{eqnarray}
This is because 
\begin{eqnarray}&&
	{\cal E}_{14}={\cal R}_{14},
\\&&
{\cal E}_{44}=({\cal R}_{44}-{\cal R}_{00}
	-{\cal R}_{11}-2{\cal R}_{22})/2.
	\label{E140=E440=0}
\end{eqnarray}
When (\ref{R00})--(\ref{R14}) are applied, 
	the last equation of (\ref{F140=F440=0})
	includes second derivatives $f_{zz}$, $h_{zz}$ and $k_{zz}$,
	while that of (\ref{E140=E440=0}) does not.


Here, we consider regions where $T_{IJ}=0$.
Then the equations in (\ref{F00=F11=F22=0}) imply the recursion formulae
\begin{eqnarray}&&
	f^{[n]}= {1 \over n(n-1)}\bigg[{f_z{}^2 \over 2f} - {f_z h_z\over 2h}
	- {f_z k_z \over k}
\cr&&\ 
	- {f_{rr} \over h}+ {f_r{}^2 \over 2fh}
	+ {f_r h_r \over 2h^2}- {f_r k_r \over kh}- {4\kappa \lambda f\over3}
	\bigg]^{[n-2]},\ \ \ \ \ 
\label{fn}
\\&&%
	h^{[n]} ={1 \over n(n-1)}\bigg[{ h_z{}^2 \over 2h}
	-{f_{z} h_{z} \over {2f}}- {h_{z} k_{z} \over {k}}
	- {f_{rr} \over f}+ {f_r{}^2 \over 2f^2}
\cr&&\ \ \ 
	+ {f_r h_r \over 2fh}
	- {2k_{rr} \over k}+ {k_r{}^2 \over k^2}+ {f_r k_r \over kh} 
	- {4\kappa \lambda h\over3}
	\bigg]^{[n-2]},\ \ \ \ \ 
\label{hn}
\\&&%
	k^{[n]}= {1 \over n(n-1)}\bigg[- {f_z k_z\over 2f}- {h_z k_z\over h}
\cr&&\ \ \ \ 
	-{ k_{rr}\over h}
	+ {f_r k_r\over 2fh}+ {h_r k_r\over 2h^2}+2- {4\kappa \lambda k\over3}
	\bigg]^{[n-2]} \ \ \ \ \ 
\label{kn}
\end{eqnarray}
for $n\ge2$.
According to (\ref{nred1})--(\ref{nred3}),
	the right-hand side of eqs.\ (\ref{fn})--(\ref{kn})
	are written in terms of
	$f^{[j]}$, $h^{[j]}$ and $k^{[j]}$ for $0\le j\le n-1$.
For example, for $n=2$,
\begin{eqnarray}&&
	f^{[2]}= {1 \over 2}\bigg({f^{[1]}{}^2 \over 2f^{[0]}} 
	- {f^{[1]} h^{[1]}\over 2h^{[0]}}
	- {f^{[1]} k^{[1]} \over r^2} 
\cr&&\ 
	- {f^{[0]}_{rr} \over h^{[0]}}
	+ {f^{[0]}_r{}^2 \over 2f^{[0]}h^{[0]}}
	+ {f^{[0]}_r h^{[0]}_r \over 2h^{[0]}{}^2}
	- {2f^{[0]}_r \over h^{[0]}r}
	- {4\kappa \lambda f^{[0]}\over3}
	\bigg),\ \ \ \ \ 
\label{f2}
\\&&%
	h^{[2]} ={1 \over 2}\bigg({ h^{[1]}{}^2 \over 2h^{[0]}}
	-{f^{[1]} h^{[1]} \over {2f^{[0]}}}- {h^{[1]} k^{[1]} \over {r^2}}
\cr&&\  
	- {f^{[0]}_{rr} \over f^{[0]}}+ {f^{[0]}_r{}^2 \over 2f^{[0]}{}^2}
	+ {f^{[0]}_r h^{[0]}_r \over 2f^{[0]}h^{[0]}}
	+ {2f^{[0]}_r \over h^{[0]}r} 
	- {4\kappa \lambda h^{[0]}\over3}
	\bigg),\ \ \ \ \ 
\label{h2}
\\&&%
	k^{[2]}= {1 \over2}\bigg(- {f^{[1]}k^{[1]}\over 2f^{[0]}}
	- {h^{[1]} k^{[1]}\over h^{[0]}}
\cr&&\ \ \ \ 
	-{ 2\over h^{[0]}}
	+ {f^{[0]}_r r\over f^{[0]}h^{[0]}}
	+ {h^{[0]}_r r\over h^{[0]}{}^2}+2
	- {4\kappa \lambda r^2\over3}
	\bigg), \ \ \ \ \ 
\label{k2}
\end{eqnarray}
where we used $k^{[0]}=r^2$ chosen using diffeomorphism.
Similarly, we can inductively determine $f^{[n]}$, $h^{[n]}$ and $k^{[n]}$
	in terms of $f^{[0]}$, $h^{[0]}$, $f^{[1]}$, $h^{[1]}$ and $k^{[1]}$ 
	for any $n>2$.
Therefore, if $f$, $h$ and $k$ and their $z$-derivatives are given at the brane,
	$f$, $h$ and $k$ off the brane
	are given in the form (\ref{calF=}) as far as it converges.
We expect that there exist regions where it converges 
	around the brane and off the core of the sphere. 
Continuations would be possible 
	by expanding the functions again around other points.
The regions where continuation fails can be taken 
	to exhibit some physical structures there.
Thus, what determine the braneworld solution are the five functions
	$f^{[0]}$, $h^{[0]}$, $f^{[1]}$, $h^{[1]}$ and $k^{[1]}$ of $r$.

\section{On-Brane Solution of the Bulk Einstein Equation }

The five functions $f^{[0]}$, $h^{[0]}$, $f^{[1]}$, $h^{[1]}$ and $k^{[1]}$
	should obey eq.\ (\ref{F140=F440=0}) which is written as
\begin{eqnarray}&&\hskip-20pt
	\frac{f^{[1]}_r}{2 f^{[0]}}
	+\frac{k^{[1]}_r}{r^2}
	=\frac{f^{[1]} f^{[0]}_r}{4 (f^{[0]})^2 }
	+\frac{h^{[1]} f^{[0]}_r}{4 h^{[0]}f^{[0]} }
	+\frac{h^{[1]} }{h^{[0]}r}
	+\frac{k^{[1]} }{r^3},\ \ 
    \label{E140=0}
\\&&\hskip-20pt 
	\frac{f^{[2]}}{ f^{[0]}}
	+\frac{h^{[2]}}{ h^{[0]}}
	+\frac{2k^{[2]}}{r^2}
	=\frac{f^{[1]}{}^2}{4 f^{[0]}{}^2 }
	+\frac{h^{[1]}{}^2}{4 h^{[0]}{}^2 }
	+\frac{k^{[1]}{}^2}{4 r^4 }
	-\frac{2\kappa\lambda}{3}.\ \ \ 
    \label{F140=0}
\end{eqnarray}
If we eliminate $f^{[2]}$, $h^{[2]}$ and $k^{[2]}$ from (\ref{F140=0})
	with (\ref{f2})--(\ref{k2}), we get 
\begin{eqnarray}
	\frac{ f^{[1]} h^{[1]} }{4f^{[0] } h^{[0]}}
	+\frac{ f^{[1]} k^{[1]} }{2f^{[0] } r^2}
	+\frac{ h^{[1]} k^{[1]} }{2h^{[0] } r^2}
	+\frac{k^{[1] }{}^2 }{4r^4}
	=\tilde R-\kappa \lambda, 
\label{E440=0}
\end{eqnarray}
where 
\begin{eqnarray}&&
	\tilde R=-\frac{1}{h^{[0]}}\bigg[
	\frac{ f^{[0]}_{rr}}{f^{[0]}}
	-\frac{ (f^{[0] }_{r})^2}{2(f^{[0]})^2}
	-\frac{ f^{[0]}_{r} h^{[0]}_{r}}{2f^{[0] } h^{[0]}}
\cr&&\hskip70pt
	+\frac{ 2f^{[0]}_{r} }{f^{[0] } r}
	-\frac{ 2h^{[0]}_{r} }{h^{[0] } r}
	-\frac{ 2h^{[0]} }{ r^2}
	+\frac{ 2}{ r^2}\bigg]\ \ \ \ \ \ 
\label{tildeR}
\end{eqnarray}
	is the scalar curvature of the brane. 
Eq.\ (\ref{E440=0}) is equivalent to ${\cal E}_{44}=0$,
	and eq.\ (\ref{E140=0}) is the same as ${\cal E}_{14}=0$
	since ${\cal R}_{14}={\cal E}_{14}$.
We have two independent differential equations 
	(\ref{E140=0}) and (\ref{E440=0})
	for five functions 
	$f^{[0]}$, $h^{[0]}$, $f^{[1]}$, $h^{[1]}$ and $k^{[1]}$.
Hence, their solution includes three arbitrary on-brane functions.
They are solvable linear differential equations as will be seen below.
Once the on-brane solution is obtained, 
	its off-brane form is always given by (\ref{calF=}) 
	with the coefficients inductively determined by (\ref{fn})--(\ref{kn}).
We assume nothing about the bulk other than the ansatz (a)--(d),
	allowing even singularities.
Therefore, the on-brane equations (\ref{E140=0}) and (\ref{E440=0})
	are essential to determine the general solution.

Let us solve the essential equations (\ref{E140=0}) and (\ref{E440=0}).
Let the functions 
\begin{eqnarray}&&
	u =- \frac {2 f^{[1]} }{f^{[0]}},\ \ 
	v =- \frac {2 h^{[1]} }{h^{[0]}},\ \ 
	w =- \frac {2 k^{[1]} }{k^{[0]}} 
\label{uvw}
\end{eqnarray}
	be arbitrary.
Then, (\ref{E140=0}) and (\ref{E440=0}) become
\begin{eqnarray}&&
	u_r+2w_r+ (u- v)f^{[0]}_r/ 2f^{[0]} +2(w-v)/r=0,
	\ \ \ \ \ \ \ 
\label{F140=01}
\\&& 
	\tilde R=2uv+4uw+4vw+2w^2+2\kappa\lambda.
\label{F440=01}
\end{eqnarray}
If $u\ne v$, eqs.\ (\ref{F140=01}) and (\ref{F440=01}) become 
	the linear differential equations
\begin{eqnarray}&&
	 f^{[0]}_r-U f^{[0]}=0, 
\label{DEforf0}
\\&& 
	(1/h^{[0]})_r+P /h^{[0]} =Q
\label{DEforh0}
\end{eqnarray}
for $ f^{[0]}$ and $1/h^{[0]}$ with
\begin{eqnarray}&&
	U= {2[-u_r-2w_r+2(v-w)/r] }/{ (u-v)},
       \label{Phi=}
\\&& 
	P= {[2U _r+(U +2/r) ^2] }/{(U +4/r)},
\label{P=}
\\&& 
	Q=\frac{-\kappa\lambda+1/r^2
	-uv-2uw-2vw-w^2 }{U/4 +1/r}.\ \ \ \ \ \ \ 
\label{Q=}
\end{eqnarray}
The solution of eqs.\ (\ref{DEforf0}) and (\ref{DEforh0}) is given by
\begin{eqnarray}&&
	f^{[0]}=e^{-\int_r^\infty U dr},\ \ \ 
       \label{f0=}
\\&& 
	h^{[0]}= e^{-\int_r^\infty P dr} 
	\left[1-\int_r^\infty Qe^{-\int_r^\infty Pdr}dr
	\right]^{-1}.\ \ \ \ 
\label{h0=}
\end{eqnarray}
If $u=v$, the functions $f^{[0]}$ becomes arbitrary, and
	eqs.\ (\ref{F140=01}) and (\ref{F440=01}) become 
	linear differential equations 
	for $w$ and $1/h^{[0]}$. 
The solution is given by
\begin{eqnarray}&&
	w=-r^{-1}\int^{\infty}_r(u-ru_r/2)dr\ \ \ 
       \label{W=}
\end{eqnarray}
and $h^{[0]} $ in (\ref{h0=}) with $P$ and $Q$ 
	obtained by the following replacements in (\ref{P=}) and (\ref{Q=}).
Replace $U$ by $ f^{[0]}_r/ f^{[0]}$
	and $w$ by the expression in (\ref{W=}).
The ranges of integration in (\ref{f0=}), (\ref{h0=}) and (\ref{W=}) 
	are chosen in accordance with the asymptotic flatness of the brane. 
The denominators in (\ref{P=}) and (\ref{Q=})
	are not identically vanishing because, if so, we have 
	$f^{[0]}= C /r^4$
	with a constant $C$,
	and the brane cannot be asymptotically flat. 
Then, the solutions for $f$, $h$ and $k$ 
	given by (\ref{calF=}) and (\ref{fn})--(\ref{kn}) with 
	$f^{[0]} $, $ h^{[0]}$, $f^{[1]} $, $ h^{[1]}$  and $ k^{[1]}$ 
	by (\ref{uvw}),	(\ref{f0=}) and (\ref{h0=})  
	expire all the solutions of the bulk Einstein equation (\ref{BE}) (alone)
	under the ansatz (a)--(d) with $ T_{IJ}=0$. 

\section{Essential Equations for Braneworld Gravity}

Now we turn to the solution for the ``braneworld".
It is a thin physical object accompanied by 
	matter distribution in the region $|z|<\delta$, 
	where $\delta$ is the infinitesimal thickness of the brane.
The expansions in $z$ in the previous chapter is possible
	in the empty regions D$^\pm$ with $\pm z>\delta$.
At $|z|<\delta$, however,
	the bulk Einstein equation with concentrated energy distribution
	indicates that $u$, $v$ and $w$ should have a gap across the brane.
We assume dominance of the collective mode $Y^I$ in $S_Y$ in (\ref{action})
	over the other terms from $S_{\rm m}$
	in the energy-momentum tensor. 
In the present coordinate system, the collective mode is given by $z=0$.
Then, the bulk Einstein equation (\ref{BE}) at $|z|<\delta$ implies
	for $\alpha=u, v$ and $w$
\begin{eqnarray}&&
	\alpha|_{z=0}=(\alpha|_{z=\delta}+\alpha|_{z=-\delta})/2
	\equiv\bar\alpha
\label{baralpha}
\\&& 
	\alpha|_{z=\delta}-\alpha|_{z=-\delta}=\kappa\tilde\lambda/3,        
\label{Delta}
\end{eqnarray}
where the last equality in (\ref{baralpha}) 
	defines $\bar u$, $\bar v$ and $\bar w$.
The Nambu-Goto equation (\ref{NG}) implies
\begin{eqnarray} 
	\bar u+\bar v+2\bar w=0.
\label{a+b+2c=0}
\end{eqnarray}
We apply the solution in the previous chapter in D$^\pm$.
At $z=\pm\delta$, 
	we have the similar equations as (\ref{F140=01}) and (\ref{F440=01}) 
	with $u^{(\pm)}$, $v^{(\pm)}$ and $w^{(\pm)}$
	in the places of $u$, $v$ and $w$, respectively.
The sums of them at $z=\pm\delta$ give
\begin{eqnarray}&&
	\bar u_r+2\bar w_r+ (\bar u-\bar v)f^{[0]}_r/ f^{[0]} 
	+2(\bar w-\bar v)/r=0,
\label{barF140=0}
\\&& 
	\tilde R=2\bar u\bar v+4\bar u\bar w +4\bar v\bar w +2\bar w^2
	+\kappa^2\tilde\lambda^2/3+2\kappa\lambda, \ \ \ \ \ \ \
\label{barF440=0}
\end{eqnarray}
while the differences trivially hold as far as we have (\ref{a+b+2c=0}).
Now, we have three equations (\ref{a+b+2c=0})--(\ref{barF440=0})
	for five functions $f^{[0]}$ and $h^{[0]}$, 
	$\bar u$, $\bar v$ and $\bar w$.
Hence, their solution includes two arbitrary on-brane functions.
They are solvable linear differential equations.
If we choose $\bar u$ and $\bar v$ arbitrarily,
	we have the same solution as 
	(\ref{f0=}), (\ref{h0=}) and (\ref{W=}) 
	with $u$, $v$, $w$ and $\lambda$
	replaced by $\bar u$, $\bar v$, $-(\bar u+\bar v)/2$
	and $\lambda+\kappa\tilde\lambda^2/6$, respectively.
Once the on-brane solution is obtained, 
	its off-brane form is always given by (\ref{calF=}) 
	with the coefficients inductively determined by (\ref{fn})--(\ref{kn}).
It expires all the solutions of our premised system.
Therefore, we conclude that
	the on-brane equations (\ref{a+b+2c=0})--(\ref{barF440=0}) 
	are essential to determine the general solution.
They are the equations which determine how the brane is curved 
	according to the dynamics, 
	and, hence, they are the essential equations for the braneworld gravity.

The general solution involve large arbitrariness,
	which may affect the predictive powers of the theory
	on the Newtonian and the post-Newtonian evidences.
It includes not only the the ordinary Schwarzschild solution on the brane,
	but also continuously deformed solutions 
	which do not satisfy the brane Einstein equation.
It implies arbitrarily deformed Newtonian potentials,
	and arbitrary amounts of light deflections and
	planetary perihelion precessions due to solar gravity.
We need further physical prescriptions to make these predictions.
For, example, conditions for the bulk behavior may improve the situation.
It is an urgent open problem of the braneworld theories.
The idea of the brane induced gravity 
	\cite {Akama82}, \cite {Akama87}, \cite{AH},
	\cite {BIG}, \cite {DvaliGabadadze}, \cite {Akama06}, 
	\cite{Gabadadze:2007dv} 
	may give a way out of this difficulty.
The brane Einstein gravity emerges through the quantum effects of the brane
	\cite{InducedGravity}.
The composite metric serves as the effective dynamical variable,
	just like the quantum-induced composite field 
	which are well understood in some classes of models \cite {ig}.

This work was supported by Grant-in-Aid for Scientific Research,
No.\ 13640297, 17500601, and 22500819
from Japanese Ministry of Education, Culture, Sports, Science and Technology.

\end{document}